\documentclass[prl,twocolumn,showpacs,superscriptaddress,nofootinbib]{revtex4}
\usepackage{graphicx}

\usepackage{bm}
\usepackage{amssymb}
\usepackage{amsmath}
\usepackage{amstext}
\usepackage{latexsym}
\usepackage[colorlinks,linkcolor=red]{hyperref}
\def\x{{\rm\bf x}}
\def\y{{\rm\bf y}}

\def\k{{\rm\bf k}}
\def\d{{\rm d}}
\def\la{\langle}
\def\ra{\rangle}
\def\om{\omega}

\newcommand{\beq}{\begin{equation}}
\newcommand{\eeq}{\end{equation}}
\newcommand{\beqa}{\begin{eqnarray}}
\newcommand{\eeqa}{\end{eqnarray}}

\begin{document}

\title{Frictionless quantum quenches in ultracold gases: a quantum dynamical microscope}

\author{A. del Campo}
\affiliation{Institut f{\"u}r Theoretische Physik, Leibniz Universit\"at Hannover, Appelstr. 2 D-30167, Hannover, Germany}
\affiliation{Institut f{\"u}r Theoretische Physik, Albert-Einstein Allee 11,
Universit{\"a}t Ulm, D-89069 Ulm, Germany}

\begin{abstract}

In this article, a method is proposed to spatially scale up a trapped ultracold gas while conserving the quantum correlations of the initial many-body state. For systems supporting self-similar dynamics, this is achieved by implementing a many-body finite-time frictionless quantum quench of the harmonic trap which acts as a quantum dynamical microscope.

\end{abstract}


\maketitle

Research in ultracold strongly correlated states of matter has recently been spurred by the experimental realization of quantum gas microscopes, allowing to detect with single site resolution and nearly unit-efficiency individual atoms within a macroscopic sample in the strongly-interacting regime optical lattice \cite{microscope1,microscope2}. These experiments are based on high-resolution optical imaging.
In a complementary way, the non-equilibrium dynamics following a quench of an external control parameter is often exploited to probe quantum correlations in  many-body systems \cite{Dziarmaga10}.
Here we propose a scheme to implement a {\it quantum dynamical microscope},
an engineered controlled expansion that allows to scale up an initial many-body state of an ultracold gas by a desired factor while preserving the quantum correlations of the initial state.
The scheme rests on the possibility of driving a self-similar dynamics in certain systems, which is a powerful tool to understand the evolution of quantum correlations. Scaling laws can often be exploited to describe harmonically trapped ultracold gases, such as the Calogero-Sutherland model \cite{Sutherland98}, the Tonks-Girardeau \cite{OS02,MG05,delcampo08} and certain Lieb-Liniger states \cite{BPG08}, Bose-Einstein condensates (BEC) \cite{CD96,KSS96,OS02}, including dipolar interactions \cite{dipolar}, strongly interacting ultracold gas mixtures \cite{GM07}, and more general many-body quantum systems \cite{Demler}.
Moreover, whenever the dynamics is not self-similar per se, it can often be assisted by tuning the interactions, either by means of Feschbach or confinement induced resonances, or time-modulation of the transverse confinement in effectively low-dimensional Bose-Einstein condensates \cite{BDZ08}. 
Nonetheless, in spite of the scaling laws, the expansion dynamics in these systems generally induces undamped breathing of the cloud
and distorts the quantum correlations of the initial state \cite{OS02,MG05,delcampo08,BPG08,CD96,KSS96,dipolar,GM07,Demler}.   
The method proposed in this Letter suppresses these effects in a finite-time non-adiabatic expansion 
which acts as a lens to zoom-up the initial state of the system.
This is achieved by carefully engineering the time-modulation of the trapping frequency 
to induce a frictionless dynamics free from  adiabaticity constraints.


{\it Self-similar dynamics}- Let us consider a $D$-dimensional many-body system composed of $N$ indistinguishable particles confined in a time-dependent harmonic trap, described by a Hamiltonian
\beqa
\label{Hamiltonian}
\hat{\mathcal{H}}\!=\!\sum_{i=1}^{N}\!\bigg[\!-\frac{\hbar^2}{2m}\Delta_i^{(D)}+\frac{1}{2}m\om^2(t)\x_i^2\bigg]\!+\!\epsilon\sum_{i<j}{\rm V}(\x_{ij}),
\eeqa
where  $\x_i\in\mathbb{R}^D$, $\x_{ij}=\x_i-\x_j$, $\Delta_i^{(D)}$ is the $D$ dimensional Laplacian operator, and $\epsilon=\epsilon(t)$ is a dimensionless time-dependent coupling constant which reduces to the identity at $t=0$.
We shall focus on systems with an interaction potential satisfying
the relation
\beqa
{\rm V}(\lambda\x)=\lambda^{\alpha}{\rm V}(\x)
\eeqa
under scaling of the coordinates.
An equilibrium state $\Phi$ of the system (\ref{Hamiltonian}) at $t=0$,  follows a self-similar evolution
\beqa
\label{scaling}
\Phi\left(\{\x_i\},t\right)=\frac{1}{b^{D/2}}e^{i\sum_{i=1}^N\!\!\frac{ m\x_i^2\dot{b}}{2b\hbar}-i\mu\tau(t)/\hbar}
\Phi\left(\!\{\frac{\x_i}{b}\},0\!\right)\!,
\eeqa
where $\mu$ is the chemical potential, and $\tau(t)=\int_{0}^tdt'/b^2(t')$, whenever the scaling factor $b=b(t)$ is the  solution of the Ermakov differential equation
\beqa
\label{EPE}
\ddot{b}+\om^2(t)b=\om_0^2/b^3
\eeqa
 with $\om_0=\om(0)$, satisfying the boundary conditions $b(0)=1$ and $\dot{b}(0)=0$.
This dynamics further requires the coupling constant and the scaling factor to be related by $\epsilon(t)=b^{\alpha -2}$, which leads to the following cases:
a) $\alpha=2$, $\epsilon(t)=1$, with no need for auxiliary tuning of the interactions, as it happens for example in a quasi-1D Bose-Einstein condensate (BEC) in the Thomas-Fermi limit or a 2D Bose gas with contact interactions \cite{KSS96,Muga}, which has recently been pointed out as an instance of a quantum anomaly \cite{Olshanii}.
b) $\alpha\neq 2$, $\epsilon(t)=b^{\alpha-2}$ which does require external tuning of the interactions as in 1D and 3D BEC to assist the self-similar dynamics
 (note that in a fast expansion the role of interactions might be disregarded, as usually done in time-of-flight experiments, and then the self-similar dynamics comes for free).
If the initial state $\Phi$ is not in equilibrium, it will then follow the evolution as if the trapping potential is kept constant in the scaled coordinates and picking the overall phase in Eq. (\ref{scaling}).
Scaling laws manifest in non-local correlations of the  gas such as
the one-body reduced density matrix (OBRDM) given by $ g_1(\x,\y;t) = \!N \!\int \!d\x_2\dots d\x_N
\Phi^*(\x,\x_2,\dots,\x_N;t)\Phi(\y,\x_2,\dots,\x_N;t)$, whose time evolution under self-similar dynamics can be conveniently written as \cite{MG05,Demler}
\begin{equation}
  \label{eq:g1_t}
  g_1(\x,\y;t) = \frac{1}{b^D} g_1\left(\frac{\x}{b},\frac{\y}{b};0\right)
  \exp\left(-\frac{i}{b}\frac{\dot{b}}{\omega_0}
  \;\frac{\x^2-\y^2}{2l^2_0}\right)
\end{equation}
where $l_0=\sqrt{\hbar/m\om_0}$, and its Fourier transform $n(\k,t) = \int d\x d\y
  \,e^{i\k\cdot(\x-\y)} g_1 (\x,\y;t)$, the momentum distribution
\begin{eqnarray}
  \label{eq:momentum_t}
  n(\k,t) &=& b^D\int \!d\x d\y\;g_1(\x,\y;0) \nonumber \\&\times&
  \exp\left[-ib\left(\frac{\dot{b}}{\omega_0}\frac{\x^2-\y^2}{2l_0^2}
  -\k\cdot(\x-\y)\right)\right],
\end{eqnarray}
as well as any higher-order correlation function, i.e.
the $n$-body reduced density matrix
$g_n(\{\x_i\}_{i=1}^n;\{\x_i'\}_{i=1}^n;t)=\frac{N!}{(N-n)!}\!\int \!
\prod_{i=n+1}^N\!\d x_i
\Phi^*(\{\x_i\}_{i=1}^N)\Phi(\{\x_i'\}_{i=1}^n,\!\{\x_i\}_{i=n+1}^N;t)
=b^{-nD}
\exp\left(-\frac{i}{b}\frac{\dot{b}}{\omega_0}
  \;\frac{\sum_{i=1}^n(\x_i^2-\x_i'^2)}{2l^2_0}\right)
  g_n(\{\frac{\x_i}{b}\}_{i=1}^n;\{\frac{\x_i'}{b}\}_{i=1}^n; 0)$.

The Newton cradle experiment has singled out the 1D Bose gas in the strongly interacting limit as a paradigmatic example of an integrable system where relaxation to equilibrium is suppressed even when perturbed to breakdown integrability \cite{Kinoshita06}.
We shall illustrate our results with a 1D cloud of ultracold bosons in the limit of hard-core contact interactions,  a Tonks-Girardeau (TG) gas, confined in a harmonic trap of frequency $\om_0$, with single-particle eigenstates $\phi_{n}(x)$, and $n=0,1,2,\dots$ The many-body ground state of this system can be elegantly described using an auxiliary wavefunction,
 a normalized Slater determinant \cite{GW00},
$\Psi_{F}(x_{1},\dots,x_{N}) =\frac{1}{\sqrt{N!}}{\rm det}_{n,l=(0,1)}^{(N-1,N)}\phi_{n}(x_{l})$,
describing a spin polarized Fermi gas in the ground-state of the trap.
This wavefunction already includes the hard-core condition as a result of the Pauli exclusion principle encoded in the determinant structure.
The bosonic symmetry can be enforced by applying the antisymmetric unit function,
$\mathcal{A}(\hat{x}_{1},\dots,\hat{x}_{N})=\prod_{1\leq j<k\leq N}\epsilon(\hat{x}_{k}-\hat{x}_{j})$, where $\epsilon(x)=1$ $(-1)$ if $x>0$ $(<0)$ and $\epsilon(0)=0$. The Bose-Fermi mapping relating both dual systems reads
$
\Phi_{TG}(x_{1},\dots,x_{N})= \mathcal{A}(\hat{x}_{1},\dots,\hat{x}_{N})\Psi_{F}(x_{1},\dots,x_{N}).
$
This is a highly non-local mapping, but being involutive, it leaves invariant any local correlation function such as the density profile, i.e. those quantities derived from the probability density are shared by both dual systems.
The computation of non-local correlations remains a non-trivial task, but
elegant expressions are known after the work by Pezer and Buljan \cite{Hrvoje}. A many-body state in the TG regime obeys the self-similar scaling law in Eq. (\ref{scaling}) whenever $b$ is a solution of Eq. (\ref{EPE}) \cite{note}.
The scaling is clearly exhibited by the density profile of a TG gas $n_{TG}(x,t)=\!N \!\int \!dx_2\dots dx_N
|\Phi_{TG}(x,x_2,\dots,x_N;t)|^2=\!N \!\int \!dx_2\dots dx_N
|\Psi_{F}(x,x_2,\dots,x_N;t)|^2=\frac{1}{b(t)}n_{TG}(\frac{x}{b(t)},0)$, its width being governed by $b$.

%
\begin{figure}[t]
\begin{center}
\includegraphics[width=0.7\linewidth]{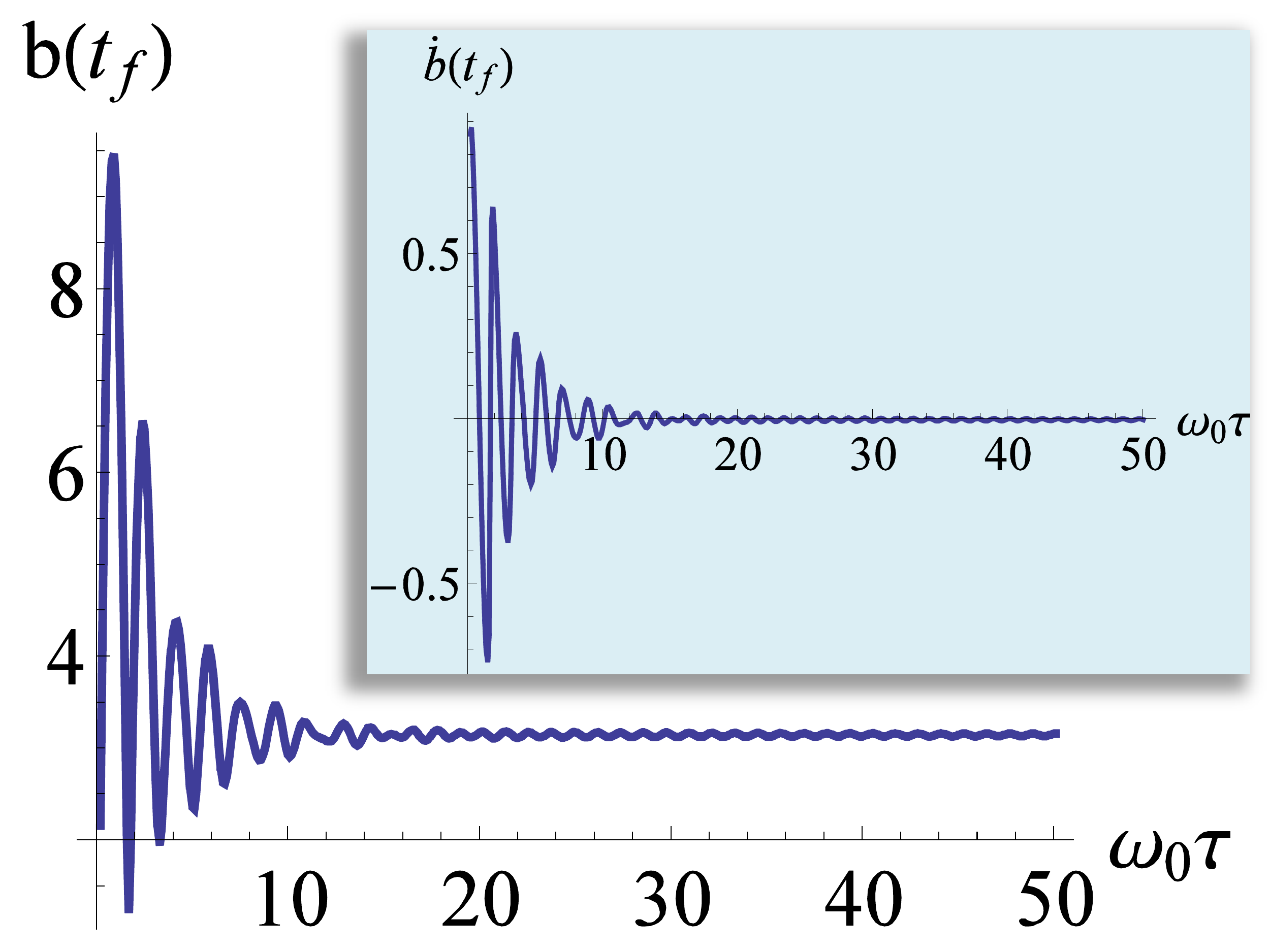}
\end{center}
\caption{\label{figbdbtanh}  Adiabatic Limit. The scaling factor following a quench for increasing quench time $\tau$ eventually approaches the adiabatic result $\gamma=\sqrt{\om_0/\om_f)}=\sqrt{10}$.
The inset shows how $\dot{b}(t_f)$ vanishes for large values of $\tau$ ($t_f=20\tau$).
}
\end{figure}

{\it Breathing and correlation dynamics}- Consider the self-similar expansion of a many-body system after quenching the trapping potential, between an initial $\om_0$ and final $\om_f$ frequency in a finite  quench time $\tau$, to scale it up by a factor $\gamma=(\om_0/\om_f)^{1/2}$. Two main features of the dynamics are to be tailored to create a quantum dynamical microscope: the subsequent breathing of the density profile and the evolution of
quantum correlations $g_n$. For illustrative purposes, we shall focus on those associated with the OBRDM.
First, we note that for the sudden expansion in free space ($\om(t>0)=\om_f=0$), $b(t)=\sqrt{1+\om_0^2t^2}$ and for
$t\gg\om_0^{-1}$, $b(t)\sim\om_0t$, $\dot{b}=\om_0$. Using the method of the stationary phase,
$n(\k,t)\sim |2\pi\om_0\l_0^2/\dot{b}|^Dg_1(\om_0 \k\l_0^2/\dot{b},\om_0 \k\l_0^2/\dot{b})$,
so the asymptotic momentum distribution is the scaled density profile of the initial state \cite{Hrvoje,Demler}, which in the TG regime can be further related to the momentum distribution of the dual system, the ideal Fermi gas \cite{RM05,MG05,delcampo08,GP08}.
This mapping between local and non-local correlations is expected as the density decreases during expansion at a finite rate $\dot{b}$.
Indeed, within the scheme of symplectic tomography, the evolution under quadratic Hamiltonians leads to a dynamical covering of correlations in phase-space \cite{DMM08}.
In an isolated system, a sudden quench of the trapping frequency between two given finite values also induces undamped breathing of the density profile. In the TG regime, this was shown by Minguzzi and Gangardt \cite{MG05}, who further illustrated the dynamics of the momentum distribution, periodically oscillating between that of the initial trapped state and the quasi-momentum distribution. Nonetheless, these two effects are ubiquitous in the family of systems supporting a dynamical scaling governed by the Ermakov equation.
For a non-vanishing $\om_f\neq 0$, increasing finite values of the quench time lead to  a gradual suppression of the subsequent breathing, and one expects to recover the adiabatic limit whenever  $\dot{\om}(t)/\om^2(t)\ll 1$.
As a specific example let us consider the functional dependence of the trapping frequency to be given by
\beqa
\label{freq_quench}
\om(t)= \om_0 +\left(\om_f-\om_0\right)\tanh\left(\frac{t}{\tau}\right)
\eeqa
with a characteristic quench time $\tau$.
The resulting $b(t)$ can be obtained by solving numerically the Ermakov equation subjected to the initial conditions $b(0)=1$, $\dot{b}(0)=0$.
The effect of an increasing value of $\tau$ on the scaling factor $b(t)$ is shown in Fig. \ref{figbdbtanh}.
%
\begin{figure}[t]
\begin{center}
\includegraphics[width=0.48\linewidth]{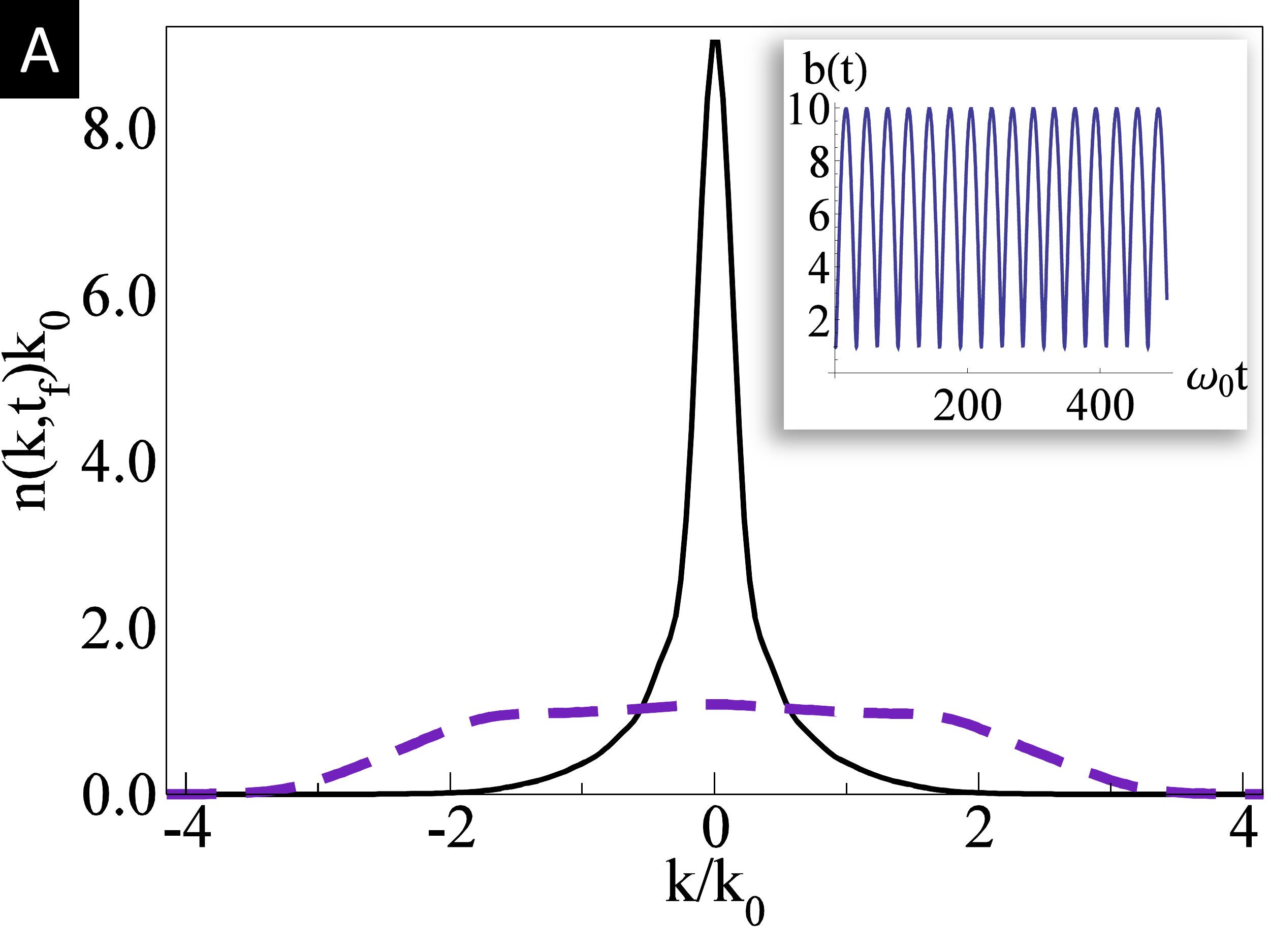}
\includegraphics[width=0.48\linewidth]{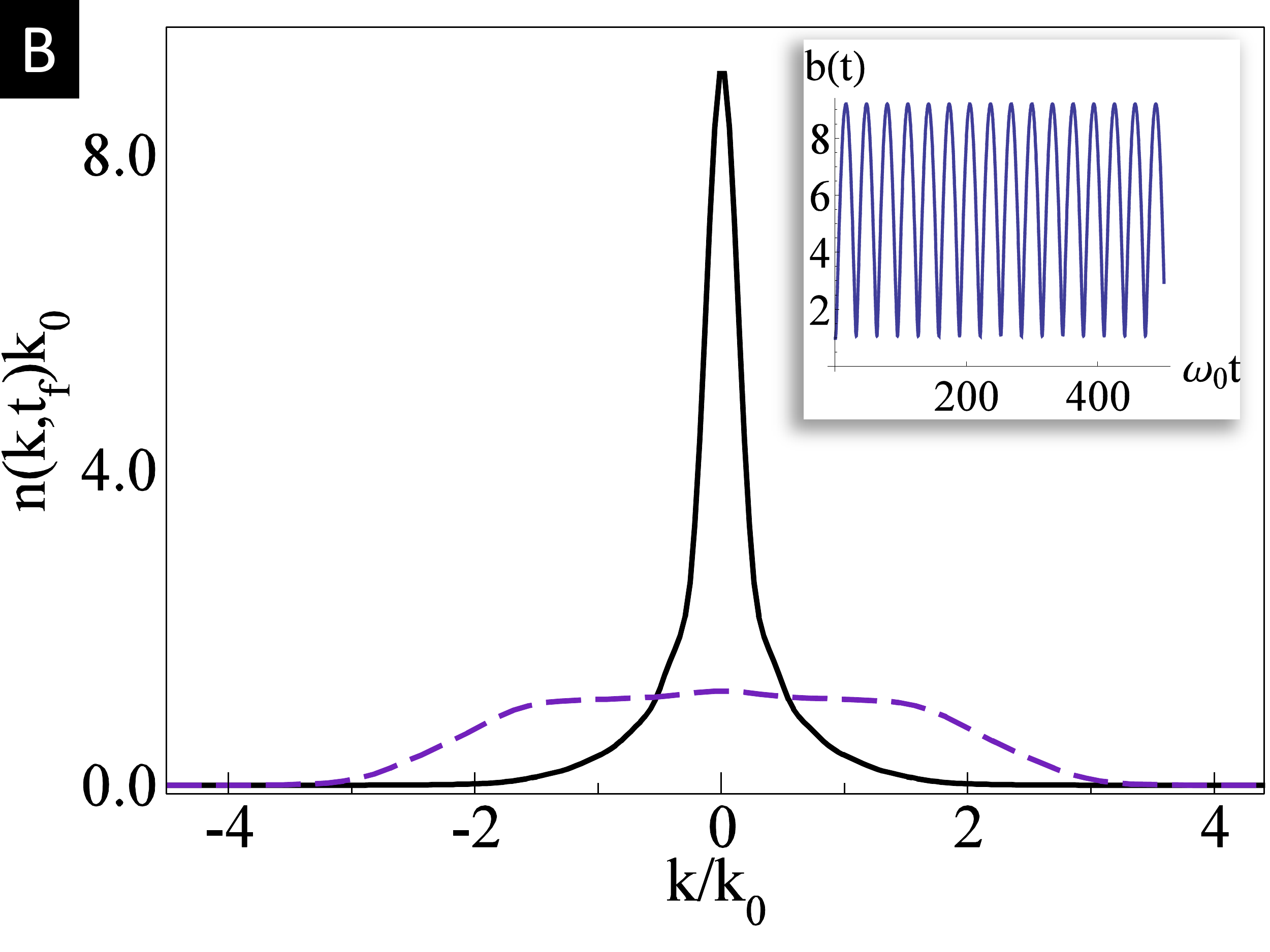}
\includegraphics[width=0.48\linewidth]{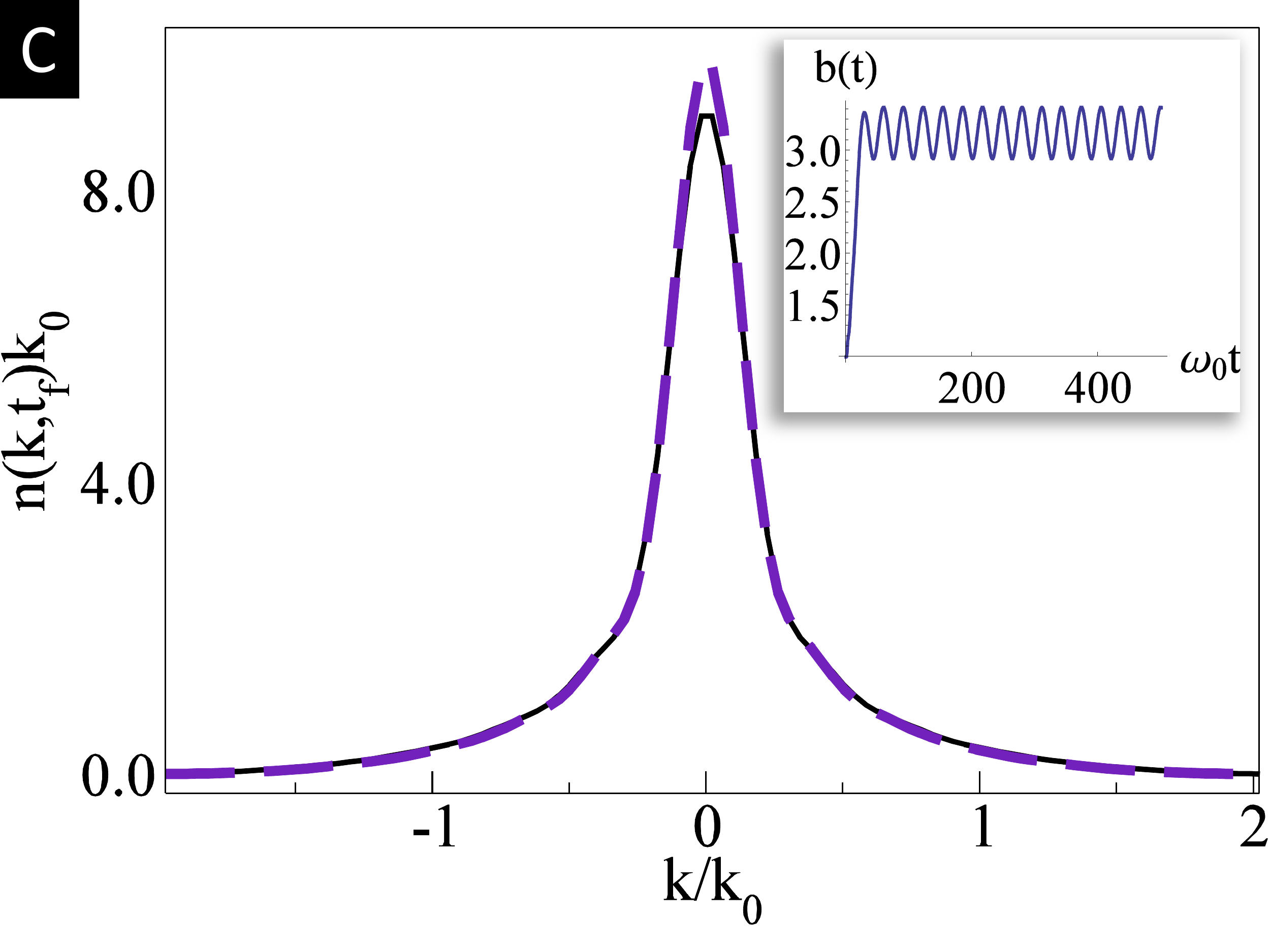}
\includegraphics[width=0.48\linewidth]{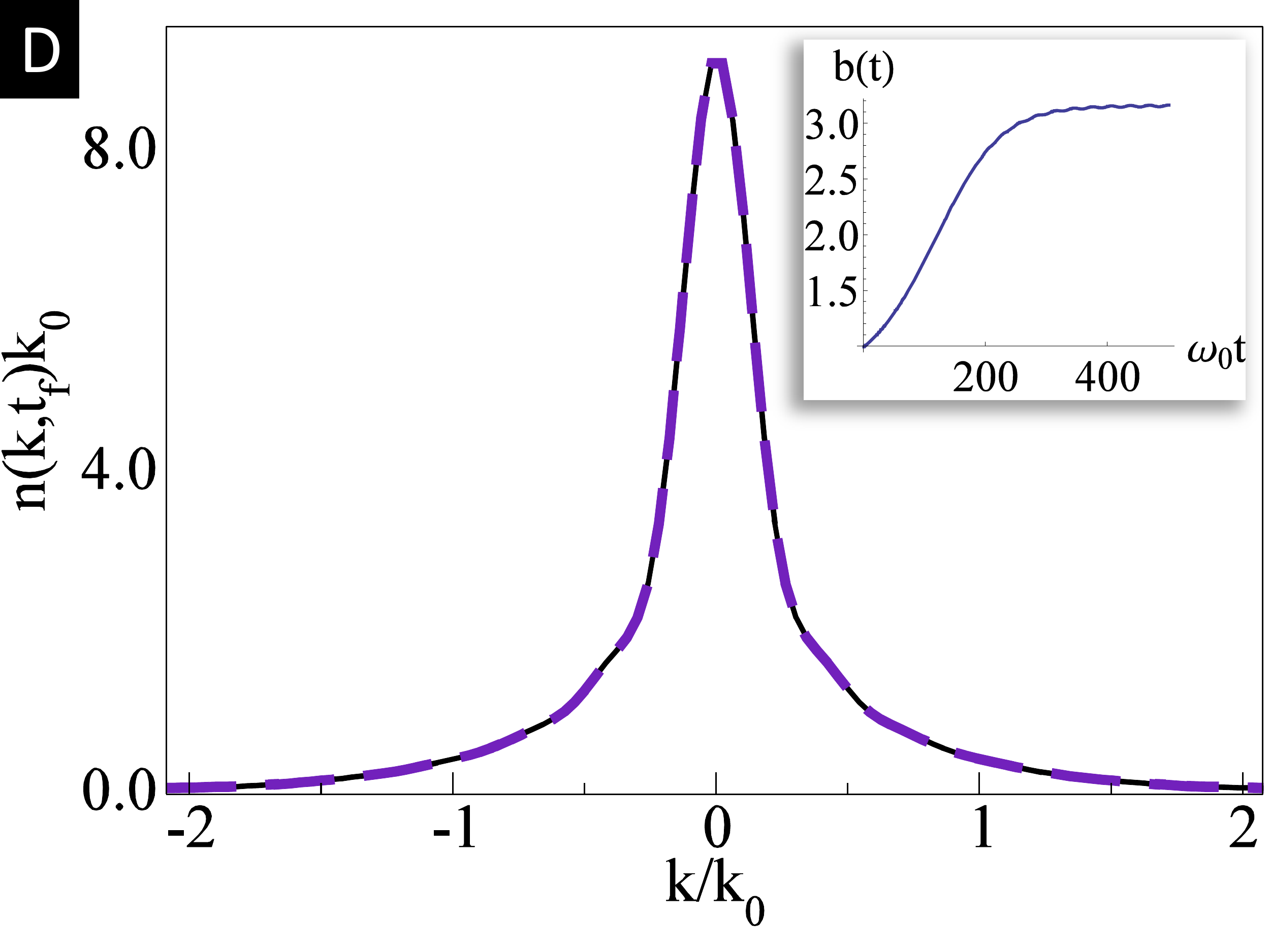}
\end{center}
\caption{\label{adiablimit}  Adiabatic Limit. Momentum distribution of a Tonks Girardeau gas following a quench of the trapping potential [see Eq. (\ref{freq_quench})] with $N=10$, $\om(0)=\om_0$ and $\om(\tau)=\om_0/10$ for increasing quench times $\tau=$: A) 0.1, B) 1, C) 10, D) 100 in units of $\om_0^{-1}$. The dashed line represents the distribution following the quench, while the solid line corresponds to the desired adiabatic limit ($t_f=500$, $k_0=\sqrt{m\om_0/\hbar}$). The insets show the evolution of the scaling function $b(t)$ along and after the quench. The adiabatic limit is approached for increasing values of $\tau\sim 100\om_0$ leading to a suppression of the breathing dynamics of the cloud and scaling of quantum correlations.
}
\end{figure}
Quenches of the potential in a finite time $\tau\sim \om_0^{-1}$ distort the momentum distribution and induce breathing of the cloud. As it turns out, these quenches do lead to non-zero finite values of $\dot{b}(t_f\gg\tau)$ and the scaling factor $b(t_f)$ deviates from the desired target value $\gamma$, as shown in Fig. \ref{adiablimit}. As $\tau$ is increased and the adiabatic limit is approached, the variation of the scaling factor becomes slower and slower, and $\dot{b}(t_f)\rightarrow 0$. Eventually, the quench follows the {\it adiabatic trajectory} $b(t)=\sqrt{\om(t)}$, for which the OBRDM  reads
\begin{equation}
  \label{scaled_g1}
  g_1(\x,\y;t) = \frac{1}{b^D(t)} g_1\left(\frac{\x}{b(t)},\frac{\y}{b(t)};0\right),
\end{equation}
this is, it is the desired result of scaling the initial OBRDM.
Similarly, the momentum distribution evolves according to
\begin{eqnarray}
  \label{scaled_np}
  n(\k,t) &=& b^D(t)\int \!d\x d\y\;g_1(\x,\y;0)
  \exp\left[i
  \gamma \k\cdot(\x-\y)\right]\nonumber\\
&=& b^D(t) n(b(t) \k,0),
\end{eqnarray}
and coincides with the initial momentum distribution up to the scaling factor $b(t)$.
Note that these expressions can be applied both for expansions ($b(t)>1$) as well as compressions ($b(t)<1$).
Nonetheless, the required adiabatic time scale can be exceedingly long and we next tackle the problem of achieving a final scaled state in a predetermined time of expansion $\tau$.


{\it Frictionless dynamics: preserving quantum correlations during expansion}-
Following the theoretical proposals in \cite{chen10,Muga} shortcuts to adiabaticity have been implemented in the laboratory both for thermal gases and  Bose-Einstein condensates \cite{Labeyrie10,Labeyrie10b}.
In the following, we show how shortcuts to adiabaticity can be exploited to control the dynamics of quantum correlations in the family of many-body systems given by Eq. (\ref{Hamiltonian}). First, we notice that the Ermakov equation can be inverted to design a many-body fast frictionless trajectory between and initial and a final trap in a given quench time $\tau$ (a time analogue of the focal plane in an optical microscope), hence, providing a shortcut to adiabaticity.
To this aim we enforce the scaling law to reduce to the initial state $\Phi$ at $t=0$ and its scaled-up form at $t=\tau$. This leads to the following boundary conditions for the scaling function, $b(0)=1$, $\dot{b}(0)=0$, $\ddot{b}(0)=0$,
$b(\tau)=\gamma=[\omega_0/\omega_f]^{1/2}$ being the scaling factor,
 and $\dot{b}(\tau)=0$, $\ddot{b}(\tau)=0$.
This set of conditions can be used to fully determine the coefficients in an ansatz, say, of polynomial type
$b(t)=\sum_{j=0}^5 a_j t^j$. One finds that
\beqa
\label{b(t)}
b (t) =
6 \left(\gamma -1\right) s^5
-15 \left(\gamma-1\right) s^4 +10 \left(\gamma-1\right)s^3
+ 1,\nonumber\\
\eeqa
with $s=t/\tau$, is a solution of the Ermakov equation that drives the evolution from the initial trapped state $\Phi(\x,0)$ to the final state $\Phi(\x,t)=\gamma^{-D/2}\Phi(\x/\gamma,t=0)$ in a finite time $\tau$
mimicking the adiabatic evolution.
%
\begin{figure}[t]
\begin{center}
\includegraphics[width=0.8\linewidth]{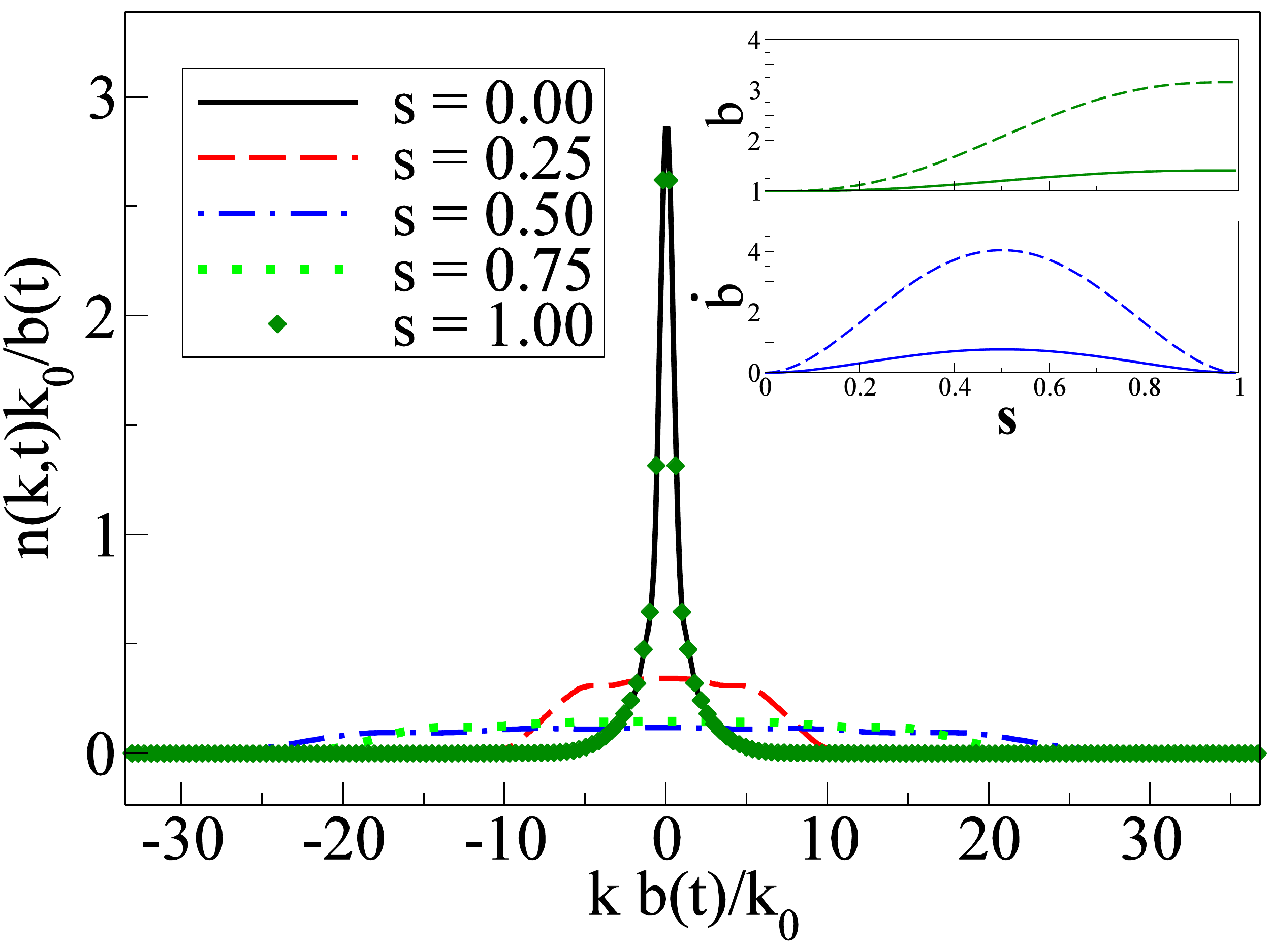}
\end{center}
\caption{\label{ffd} (color online). Frictionless quantum quench of a Tonks-Girardeau gas.
Evolution of the momentum distribution exhibiting signs of dynamical fermionization at an intermediate stage of the expansion, before reaching the final time $\tau$ at which it reduces to the scaled-up distribution of the initial trapped state ($\gamma=10$, $N=10$). The inset shows the smooth evolution of the scaling factor in a frictionless expansion in an arbitrarily short quench time $\tau$, for both  $\gamma=\sqrt{2}$ (solid line) and $10$ (dashed line) expansion factors.
}
\end{figure}
The required modulation of the trapping frequency can be obtained as well from Eqs. (\ref{EPE}) and (\ref{b(t)}), and might involve imaginary frequencies associated with a repulsive potential $\om^2<0$ \cite{chen10} whenever the demanded expansion time  is small, $\tau\lesssim\om_0^{-1}$. Should that be required, the trajectory can be implemented in the laboratory as in \cite{Khaykovich}.
The upshot of the frictionless dynamics is that quantum correlations at the end of the quench ($t=\tau$, and only then) are those of the initial state scaled by a factor $b(\tau)=\gamma$. In particular,
\beqa
\label{QDMcorr}
g_1(\x,\y;\tau) &=& \frac{1}{\gamma^D} g_1\left(\frac{\x}{\gamma},\frac{\y}{\gamma};0\right),\nonumber\\
n(\k,\tau) &=& \gamma^D n(\gamma \k,0).
\eeqa
Similar expressions hold for higher-order correlations, i.e.
$g_n(\{\x_i\},\{\y_i\};\tau) = \gamma^{nD} g_n\left(\{\x_i/\gamma\},\{\y_i/\gamma\};0\right)$, with $\{\x_i\}=\{\x_i\}_{i=1}^n$.
Moreover, as long as the initial state is an equilibrium state in the initial trap, so it is the state at $\tau$ with respect to the final trap, preventing any non-trivial dynamics after the quench, for $t>\tau$ if $\om(t>\tau)=\om_f$.
Nonetheless, at intermediate times $t\in [0,\tau)$ the momentum distribution exhibits a rich non-equilibrium dynamics, and can show for instance, evolution towards the scaled  density profile of the initial state.
Note that this mapping is favored by a large scaling factor $\gamma$.
%

Implementing a frictionless dynamics allows to perform a controlled expansion towards the scaled-up state, as illustrated in Fig. \ref{ffd}.
For a given final time $\tau$ of expansion, the momentum distribution approaches most closely the initial density around $t\approx \tau/2$. As shown in the inset,  $\dot{b}$ has a maximum precisely at $t=\tau/2$ around which it is approximately constant as in the asymptotic free expansion. For $t>\tau/2$ the evolution proceeds so as to reconstruct the initial state scaling it by the desired factor $\gamma$.
We close noticing that the process is robust in the sense that within linear response, 
errors in the implementation of the trap modulation or a many-body perturbation represented by the operator $\varepsilon {\rm W}(\x,t)$ (see the supplementary material) lead only to a quadratic decay of the fidelity between the target state $\Phi(\tau)$ and the resulting state at the end of the quench $\Phi'(\tau)$, $\mathcal{F}(\tau)=|\la\Phi(\tau)|\Phi'(\tau)\ra|^2 =1-\varepsilon^2(\tau/\tau_Z)^2+\mathcal{O}(\varepsilon^4)$, where the Zeno-like time $\tau_Z=\hbar/[\la \widetilde{{\rm W}}^2\ra-\la \widetilde{{\rm W}}\ra^2]^{1/2}$, 
with $\widetilde{{\rm W}}=\frac{1}{\tau}\int_0^{\tau}U_0(-t){\rm W}(t)U_0(t)]dt$ where $U_0$ is the time evolution operator. Hence, the effect of $\varepsilon {\rm W}(\x,t)$ is controlled by the ratio $\tau/\tau_Z$ and can be suppressed in fast expansions.

In conclusion, for many-body systems supporting self-similar dynamics, we have shown how to scale up the system by means of a controlled expansion without modifying the quantum correlations. As an alternative to the adiabatic dynamics, we propose to implement a fast frictionless quench of the trapping potential, which acts as a quantum dynamical microscope, leading to the scaled-up initial state at the end of the quench.

Insightful discussions with  J. G. Muga, M. B. Plenio and L. Santos are acknowledged, as well as financial support by EPSRC and the European Commission (HIP).

\section{Supplementary material}

\section{Stability against deviations from the designed trap modulation}

Using the Ermakov equation Eq. (\ref{EPE}),and the scaling factor $b(t)$ in Eq. (\ref{b(t)}) one can obtain the trajectory  
\begin{widetext}
\beqa
\frac{\om^2(t)}{\om_0^2}=\frac{\om_0^2\tau^2-60 (\gamma-1) s (s (2 s-3)+1) \left((\gamma-1) (3 s (2 s-5)+10) s^3+1\right)^3}{\left((\gamma-1) (3 s (2 s-5)+10) s^3+1\right)^4 \om_0^2\tau^2},
\eeqa
\end{widetext}
 with $s=t/\tau$, 
which leads to a successful implementation of the quantum dynamical microscope. 
Clearly, the pair $\{b(t),\om(t)\}$) is not unique, and provided that $b(t)$ satisfies the boundary conditions at $t=0,\tau$, many 
alternatives can be found.
In practice, deviations of these trajectories are likely to occur, 
and we next consider the stability against a time-dependent frequency shift $\delta(t)$ associated 
with the one-body perturbation $\varepsilon{\rm W}(\x,t)=\varepsilon\sum_iv(x_i,t)=\varepsilon\sum_i\frac{1}{2}m\delta\om^2(t)x_i^2$ ($\delta\om^2(t)=2\om(t)\delta(t)+\delta^2(t)$).
As an illustrative example, let us consider the fidelity $\mathcal{F}(\tau)=|f(\tau)|^2$ of a N-body state in the TG regime 
which at $t=0$ is the ground state of the trap, and where $f(\tau)=\la\Phi(\tau)|\Phi'(\tau)\ra$ is the fidelity amplitude between the target $\Phi(\tau)$ and perturbed $\Phi'(\tau)$ states. 

In an expansion on the strength of the perturbation, the Zeno-like time governing the decay of 
$\mathcal{F}(\tau)=1-\varepsilon^2(\tau/\tau_Z)^2+\mathcal{O}(\varepsilon^4)$ 
is found to be ($N>1$)
\beqa
\tau_Z=\frac{2\sqrt{2}\om_0\tau}{N\int_0^{\tau}dt\delta\om^2(t)b^2(t)}. 
\eeqa

According to Eq.(\ref{b(t)}), $b(t)\leq\gamma$ for all $t\in[0,\tau]$, so that a high-fidelity microscope requires
\beqa
\label{boundtau}
\tau\ll\frac{2\sqrt{2}\om_0}{N\gamma^2\overline{\delta\om^2}}.
\eeqa
Hence, the fidelity decay is governed by the time average of the square of the frequency shift $\overline{\delta\om^2}=\frac{1}{\tau}\int_0^{\tau}dt\delta\om^2(t)$. In particular, for a Gaussian white-noise stochastic variable $\delta(t)$ of zero mean and covariance $\delta_0^2$,  
averaging over different realizations leads to $\la\overline{\delta\om^2}\ra=\delta_0^2$. 
We further note that the robustness against shifts in time of the ideal trajectory $\om(t)$, is warranted by the smoothness of the scaling factor $b(t)$ in the final stage of the quench, in the sense that $b(t)\simeq\gamma$, $\dot{b}(t)\simeq 0$.

\section{Stability against perturbations of the many-body Hamiltonian}

Physical implementations of the family of ultracold gases described by Eq. (\ref{Hamiltonian}) generally 
include deviations represented by a many-body perturbation $W(\x)$.
In the following, we study the robustness of the quantum dynamical microscope along a 
trajectory $b(t)$ designed ignoring a pair-wise perturbation $\varepsilon{\rm W}(\x)=\varepsilon\sum_{i<j}v(x_i-x_j)$, and compare the result  with that of the full dynamics governed by the perturbed Hamiltonian.
To illustrate this we shall consider and ultracold cloud in the Tonks-Girardeau regime 
as described in the text, for which the trapped modulation $\omega(t)$ and scaling function $b(t)$ are designed. 
Assume that the real cloud prepared in the laboratory is a strongly interacting Bose gas, 
but not strictly in the TG regime. Generally, a Lieb-Liniger Bose gas with finite interactions in a time-dependent trap, described by a many-body Hamiltonian 
\beqa
\hat{\mathcal{H}}\!=\!\sum_{i=1}^{N}\!\bigg[\!-\frac{\hbar^2}{2m}\partial_{x_i}^2+\frac{1}{2}m\om^2(t)x_i^2\bigg]
+g_{1D}^B\sum_{i<j}\delta(x_i-x_j),\nonumber
\eeqa
 does not follow a self-similar expansion under a trap modulation $\omega(t)$ unless the interaction strength is tuned simultaneously according to $g_{1D}^B(t)=g_{1D}^B/b(t)$ as discussed in the main text, case b) with $\alpha=1$.
The effect of finite interactions can be taking into account by considering the perturbation on the TG dynamics represented by 
the two-body $\delta''$-pseudopotential \cite{Sen}, 
\beqa
U(\x)=-\frac{2\hbar^4}{m^2g_{1D}^B}\sum_{i<j}\delta''(x_i-x_j).
\eeqa
The breakdown of self-similar dynamics induced by it can be suppressed by imposing
\beqa
\label{boundtau2}
\tau\ll\frac{ m^2g_{1D}^B}{2\hbar^3k_0^3}
\frac{\overline{b(t)^3}}{C_N},
\eeqa
where $C_N<N^2$ is a coefficient depending on the number of particles.

The upshot is that perturbations render unfeasible an adiabatic version of the quantum dynamical microscope, 
while employing an engineered finite-time expansion makes it stable against both one and two-body perturbations. 
This follows from linear response theory which to second order on the strength of the perturbation
imposes a quadratic decay of the fidelity  controllable by the quench time $\tau$ associated with the expansion.


\begin{thebibliography}{99}
\bibitem{note} Note that this scaling is exclusive of the limit with hard-core interactions, and that when the initial state is  a Lieb-Liniger gas with finite-coupling constant, there is a breakdown of the scaling law during a 1D expansion, and only in a late stage, as the decreasing density brings the system effectively into the Tonks-Girardeau regime, the dynamics becomes self-similar \cite{OS02}.
Nonetheless, it is possible to induce a self-similar dynamics on a Bose gas with finite time-dependent contact interactions whenever $\epsilon(t)=1/b(t)$.
\end{thebibliography}
\end{document}